\documentclass{appolb}
\usepackage{graphicx}
\usepackage{slashed}
\usepackage{cite}

\begin{document}
\title{Chiral-symmetry breaking and pion structure in the Covariant Spectator Theory %
\thanks{Presented at Excited QCD 2016}%
}
\author{Elmar P. Biernat, M. T. Pe\~{n}a
\address{CFTP, Instituto Superior T\'{e}cnico (IST), Universidade de Lisboa, Av. Rovisco Pais, 1049-001 Lisboa, Portugal}
\\
{Franz Gross
}
\address{Thomas Jefferson National Accelerator Facility (JLab),
Newport News, Virginia 23606, USA and College of William and Mary, Williamsburg, Virginia 23188,
USA}
\\
{Alfred Stadler
}
\address{Departamento de F\'{\i}sica, Universidade de \'{E}vora, 7000-671 \'{E}vora, Portugal}
\\
{Em\'{\i}lio Ribeiro
}
\address{CeFEMA, Instituto Superior T\'{e}cnico (IST), Universidade de Lisboa, 1049-001 Lisboa, Portugal}
}
\maketitle
\begin{abstract}

We introduce a covariant approach in Minkowski space for the description of quarks and mesons
that exhibits both chiral-symmetry breaking and confinement. In a simple model for the interquark interaction the quark mass function is obtained and used in the calculation of the pion form factor. We study the effects of the mass function and of the different quark pole contributions on the pion form factor.

\end{abstract}
  
\section{Introduction}
The high-prescision measurements at the Jefferson Lab accelerator after its 12 GeV upgrade will provide new data on the pion form factor which cover the interesting region up to momentum transfer $Q^2 \approx 6$ GeV$^2$ where the pion form factor scaled with $Q^2$ has a maximum~\cite{Dudek:2012fk}. Together with theoretical calculations, they will narrow the uncertainty about the smallest $Q^2$ at which the description based on asymptotic parton distribution functions is still valid and they will clarify the mismatch between experiment and perturbative QCD predictions in the region around $Q^2 \approx 6$ GeV$^2$~\cite{Chang:2013qy}. This will also help to resolve the current discrepancy between the results for the $\pi \gamma^* \gamma$ transition form factor obtained by the Babar and Belle Collaborations.

In this article we focus on the theoretical calculation of the pion form factor in the spacelike ($Q^2>0$) region. Here, the small-$Q^2$ region is of particular interest because of its vicinity to the timelike ($Q^2 <0$) sector. The pion form factor enters into the evaluation of baryon form factors near $Q^2 \approx 0$, and its behavior in the timelike region can help in the interpretation of dilepton production data from heavy ion collisions.

Various modern theoretical approaches have addressed the non-perturbative dynamics underlying the pion and other hadronic bound states. For instance, QCD simulations on the lattice~\cite{Edwards}, quantum field theory formualted on the light front~\cite{Brodsky:1997de}, as well as models based on the Dyson-Schwinger/Bethe-Salpeter (DSBS) approach and the mass gap equation~\cite{Bicudo:1989sh} have made significant contributions to our understanding of hadron phenomenology.

We use a framework similar to the DSBS approach, the Covariant Spectator Theory (CST) \cite{Gro69}, in which a quark mass is dynamically generated in a way consistent with the quark-antiquark dynamics by satisfying the axial-vector Ward-Takahashi identiy (AVWTI) \cite{PhysRevD.90.096008}. In contrast to the DSBS approaches the CST equations are solved in Minkowski space, which allows, for instance, a straightforward extension of pion form factor results from the spacelike to the timelike $Q^2$ region.

\section{CST model for $q\bar q$ mesons}
\begin{figure}[htb]
\centerline{%
\includegraphics[width=12.5cm]{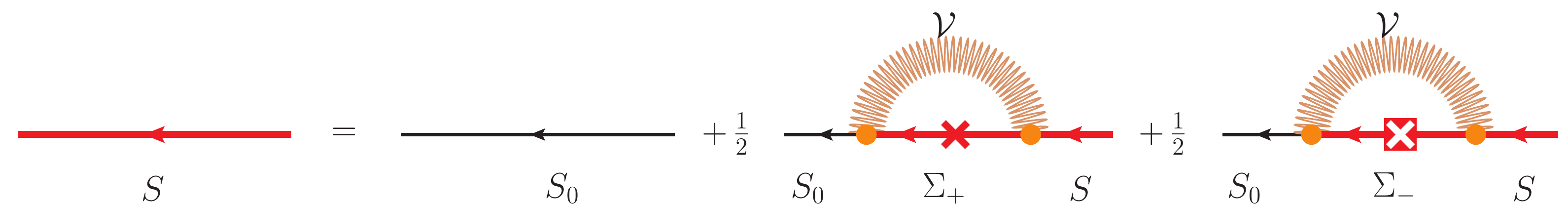}}
\caption{Diagrammatic representation of the CST-DE.  The thick arrowed red lines with and without red (white) crosses represent positive (negative) on-shell quark projectors and dressed off-shell quark propagators $S$, respectively. The thin arrowed black lines are the bare quark propagators $S_0$ and $\Sigma_+$ ($\Sigma_-$) are self-energy contributions from the positive (negative) energy quark pole. The orange zigzag line is the interaction kernel $\mathcal V$.}
\label{Fig:CSTBSE}
\end{figure}

\begin{figure}[htb]
\centerline{%
\includegraphics[width=8cm]{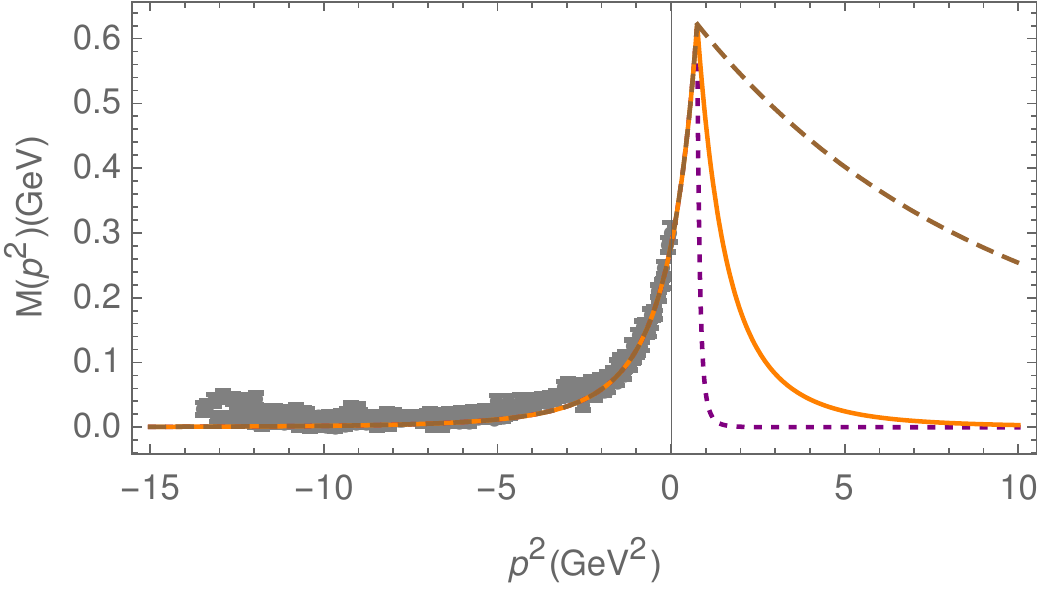}}
\caption{The chiral limit mass function with 3 different possible shapes in the timelike region compared with the lattice data~\cite{Bowman:2005vx} extrapolated to the chiral limit.}
\label{Fig:massfunction}
\end{figure}

Within the charge-conjugation invariant CST framework~\cite{Savkli:1999me,PhysRevD.89.016005} the CST-Dyson equation (CST-DE) for the dressed quark propagator is obtained from the Dyson equation by keeping only the quark propagator pole contributions in the loop four-momentum integration. Its diagrammatic representation is given in Fig.~\ref{Fig:CSTBSE}.
The CST-DE describes the dynamical generation of the quark self-energy $\Sigma(p)\equiv\Sigma_+(p)+\Sigma_-(p)=A(p^2)+\slashed pB(p^2)$ in the dressed quark propagator $S(p)=[m_0+ \Sigma(p)-\slashed p-\mathrm i\epsilon ]^{-1}$, where $m_0$ is the bare quark mass, $p$ is the off-shell quark momentum and the dynamical quark mass function is defined by 
\begin{equation}
 M(p^2)=\frac{A(p^2)+m_0}{1-B(p^2)}\,. 
\end{equation}
The constituent quark mass $m$ is defined as the value of the mass function where $S$ has a pole, i.e., $M(m^2)=m$. 
We use an interaction kernel of the form
\begin{equation}
 \mathcal V(p,\hat k)=(\mathbf{1}\otimes\mathbf{1}+\gamma^5\otimes\gamma^5) V_L(p,\hat k)+\gamma^{\mu} \otimes\gamma_\mu h^2(p) \frac{C}{2m} (2\pi)^3E_k \delta^3(\vec p-\vec k)  \label{eq:V}
\end{equation}
where $V_L(p,\hat k)$ is the CST generalization in momentum space of the linear confining potential satisfying \cite{Gro69}
\begin{eqnarray}
 \int \frac{\mathrm d^3 k}{E_k} V_{L} (p,\hat k)=0\,,\quad {\rm where}\quad E_k=\sqrt{m^2+\vec k^2}\,,\quad {\rm and}\quad  \hat k=(E_k,\vec k)\,.
 \label{eq:VLzero}
\end{eqnarray}
The second term in Eq.~(\ref{eq:V}) is the CST generalization of the constant potential where $C$ is its strength and $h(p^2)$ is a strong quark form factor. It has been shown~\cite{PhysRevD.90.096008} that the kernel $\mathcal V(p,\hat k)$, when applied in both CST-BSE and CST-DE, satisfies the AVWTI and complies with the Adler-zero constraint~\cite{Adler_PhysRev.137.B1022} in $\pi$-$\pi$-scattering imposed by chiral symmetry. This is because the linear confining term of $\mathcal V(p,\hat k)$ does not contribute to the CST-DE. Further, $B=0$ in this simple model, and the dynamical quark mass function in the chiral limit where $C=m$ assumes the form
\begin{eqnarray}
 M(p^2)=m h^2(p^2)\,. 
 \label{eq:massfunction}
\end{eqnarray}
The strong quark form factor $h(p^2)$ depends on $m$ and a cutoff parameter, which are determined by a fit of $M(p^2)$ at negative $p^2$ to the lattice QCD data~\cite{Bowman:2005vx} extrapolated to the chiral limit. In the timelike region ($p^2>0$), for which no lattice data are available, we adopt a piecewise form. Varying the shape of $h$ in this region will allow us to study the sensitivity of the pion form factor to the functional form of $h$. Figure~\ref{Fig:massfunction} shows the mass function together with the lattice data in the chiral limit.

%

\section{Triangle diagram and the pion form factor}
\begin{figure}[htb]
\centerline{%
\includegraphics[width=8cm]{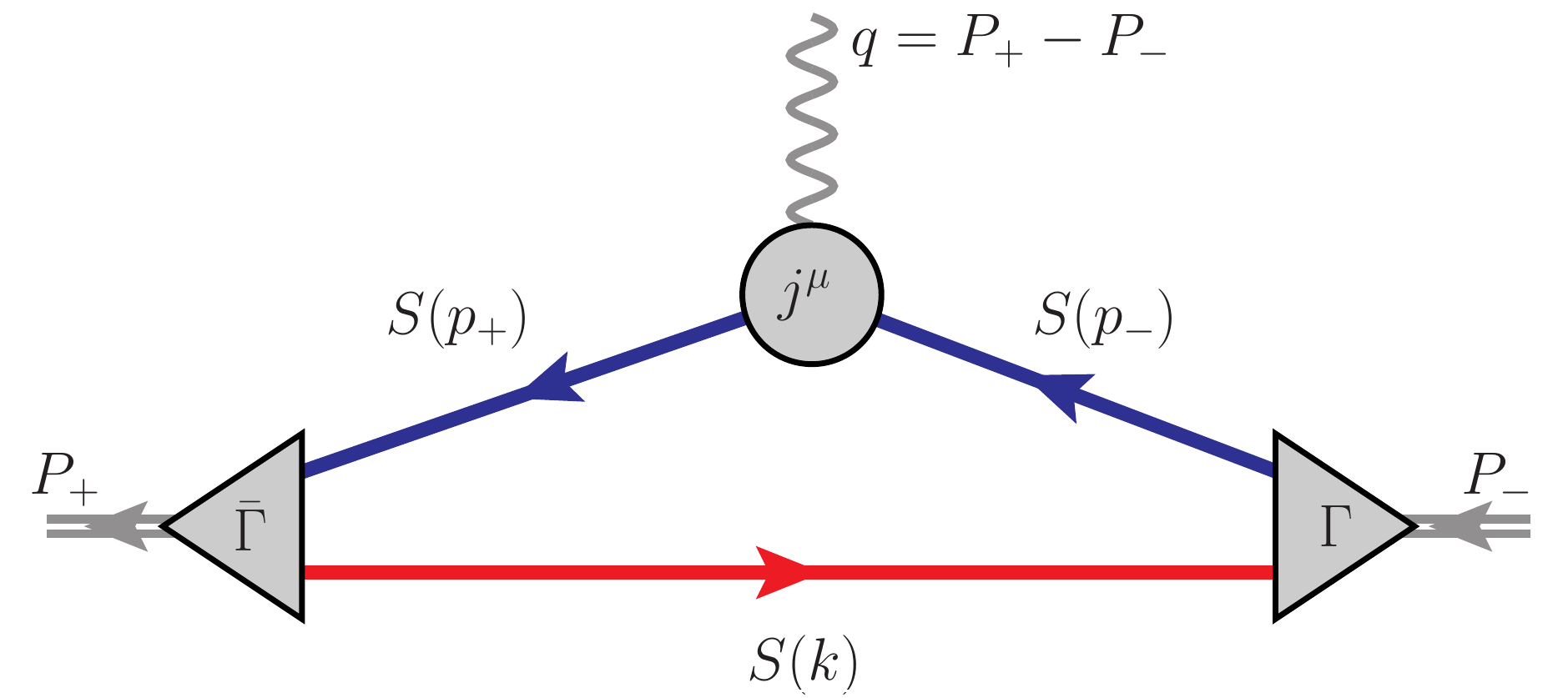}}
\caption{The triangle diagram which describes the interaction of the virtual photon with the quark (blue), with the antiquark as a spectator (red).}
\label{Fig:TriangleA}
\end{figure}

The elastic electromagnetic pion form factor is obtained in impulse approximation from the sum two triangle diagrams, in which the photon couples either to the quark or the antiquark. The first of these diagrams is depicted in Fig.~\ref{Fig:TriangleA}. 
In order to evaluate the triangle diagram using the charge-conjugation invariant CST prescription of how to perform the energy-contour integration requires taking all quark propagator-pole contributions into account, i.e. the 4 poles of the active quark at $p_+^2=m^2$ and $p_-^2=m^2$ and the 2 spectator quark poles at $k^2=m^2$~\cite{Biernat:2015xya}.

One ingredient of the pion form factor calculation is the pion
vertex function $\Gamma$. Instead of solving the full CST-BSE we use the approximated pion vertex function near the chiral limit of the form $\Gamma (p_1,p_2)\propto h(p_1^2)h(p_2^2)\gamma^5$~\cite{Biernat:2015xya}. The other ingredient is the quark current which should also be calculated from solving the inhomogeneous CST-BSE. Here we use, however, for simplicity, the current proposed in Ref.~\cite{Biernat:2015xya} which applies the framework by Riska and Gross \cite{Gro87} to ensure gauge invariance. 
\section {Results and conclusions}
\begin{figure}[htb]
\centerline{%
\includegraphics[width=7cm]{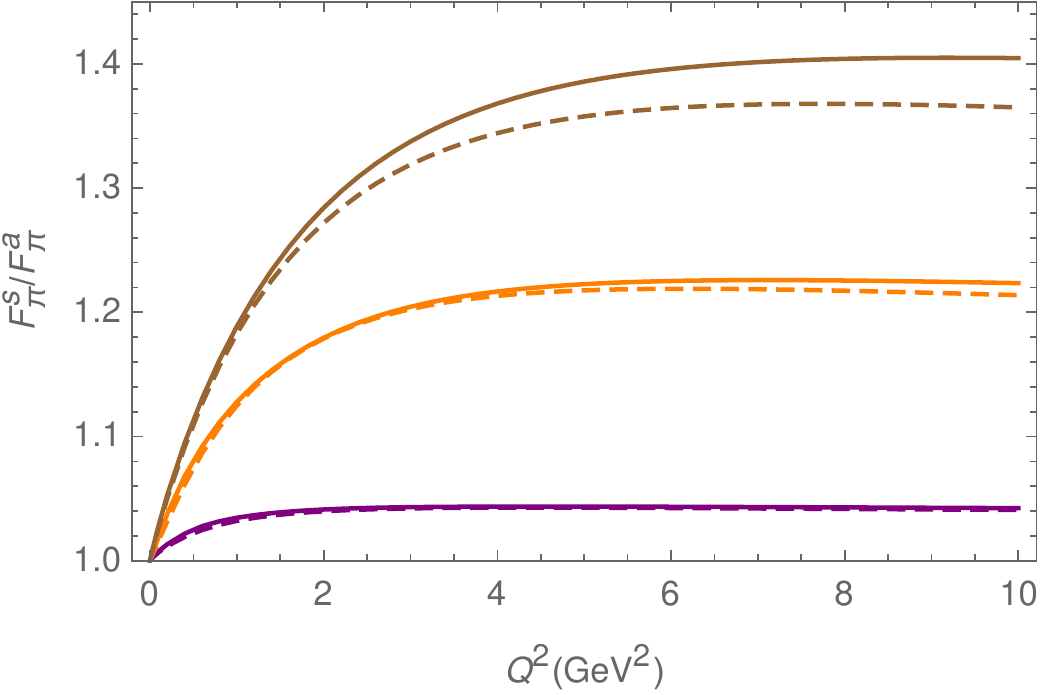}}
\caption{The ratio $F_\pi^s/F_\pi^a$ for fixed (dashed lines)
and running (solid lines) quark masses, and different values of $m_\pi$.
The pairs of curves, from top to bottom, are the results obtained
with $m_\pi=0.6$ (brown), $0.42$ (orange), and $0.14$ GeV (purple).}
\label{Fig:SpectOverAct_running_vs_fixed}
\end{figure}

In Fig.~\ref{Fig:SpectOverAct_running_vs_fixed} we present the ratio of the spectator pole contributions $F^s_\pi$ and active pole contributions $F^a_\pi$ calculated with fixed and running quark masses and different values of the pion mass $m_\pi$. 

\begin{figure}[htb]
\centerline{%
\includegraphics[width=7cm]{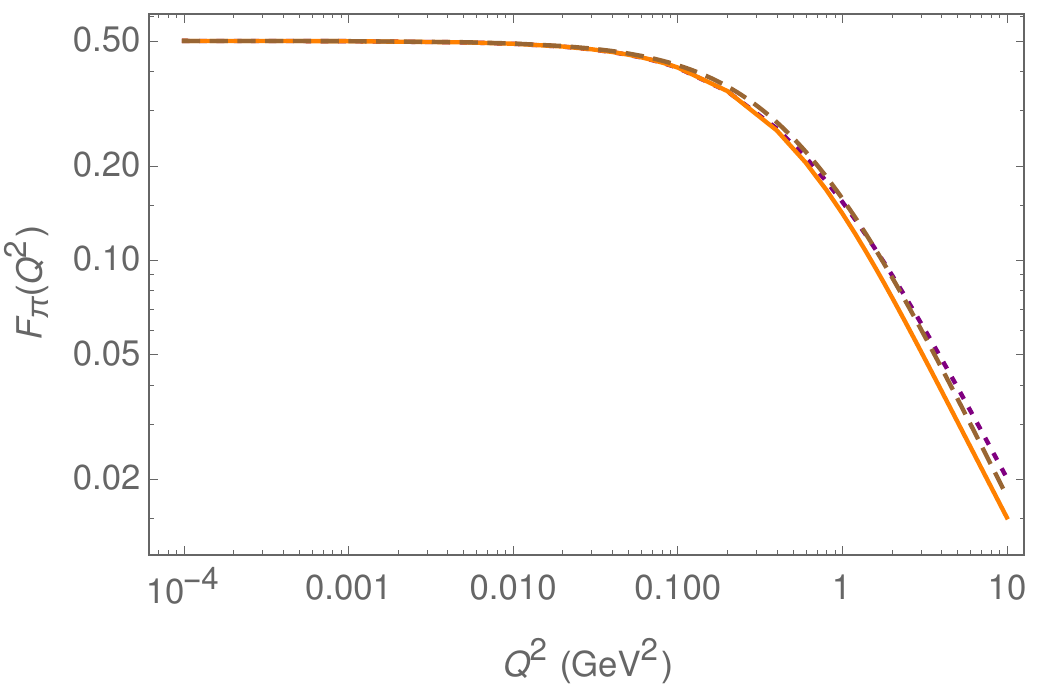}}
\caption{$F_\pi^a$ when calculated with the mass functions of Fig.~\ref{Fig:massfunction}.}
\label{Active_Poles_dynamical_mu_420_var_alpha}
\end{figure}

In Fig.~\ref{Active_Poles_dynamical_mu_420_var_alpha} we compare the results for $F_\pi^a$ when calculated with different mass functions in the timelike region of Fig.~\ref{Fig:massfunction}. 
Note that the computation of $F_\pi^s$ tests the mass function only in the spacelike region and thus all curves coincide in this case.
We conclude that for the present simple model for small $m_\pi$ the active quark contributions are as important as the spectator contributions, over the whole range of $Q^2$. For large $m_\pi$ and large $Q^2$, the active pole contributions are suppressed as compared to the spectator contributions by about 30\%.  This suppression is slightly stronger for running than for fixed quark masses. For small $m_\pi$, the spectator and active pole contributions are nearly identical, not only in magnitude but also in shape, even for large $Q^2$. Furthermore, we find that the pion form factor is surprisingly insensitive to the functional form of the strong quark form factors and quark mass function.


\subsection*{Acknowledgements}
This work was supported by Funda\c c\~ao para a Ci\^encia e a 
Tecnologia (FCT) under Grants No. CFTP-FCT (UID/FIS/00777/2013), No. CERN/FP/123580/2011, and No. SFRH/BPD/100578/2014, and by the European Community's Seventh Framework Programme FP7/2007-2013 under Grant Agreement No.\ 283286. F.G. was supported by the U.S. Department of Energy, Office of Science, Office of Nuclear Physics under contract DE-AC05-06OR23177.


\end{document}